\documentclass[conference]{IEEEtran}

\usepackage{amsmath}
\usepackage{amssymb}
\usepackage{cite}
\usepackage{epsf}
\usepackage{dsfont}
\usepackage{flushend}

\newtheorem{thm}{Theorem}

\newtheorem{col}{Corollary}

\begin{document}
\title{Precoding in Multiple-Antenna Broadcast Systems with a Probabilistic Viewpoint}
\author{Amin Mobasher and Amir K. Khandani\\
Coding \& Signal Transmission Laboratory (www.cst.uwaterloo.ca),\\
Dept. of Elec. and Comp. Eng., University of Waterloo, Waterloo,
ON, Canada, N2L 3G1,\\
E-mail: \{amin, khandani\}@cst.uwaterloo.ca
}
\maketitle

\footnotetext[1]{This work is supported by the Nortel Networks, the
Natural Sciences and Engineering Research Council of Canada (NSERC),
and the Ontario Center of Excellence (OCE).}

%
\begin{abstract}
In this paper, we investigate the minimum average transmit energy
that can be obtained in multiple antenna broadcast systems with
channel inversion technique. The achievable gain can be
significantly higher than the conventional gains that are mentioned
in methods like \cite{PeHoSw052}. In order to obtain this gain, we
introduce a Selective Mapping (SLM) technique (based on random
coding arguments). We propose to implement the SLM method by using
nested lattice codes in a trellis precoding framework.
\end{abstract}

\section{Introduction}

Recently, there has been a considerable interest in Multi-Input
Multi-Output (MIMO) antenna systems due to achieving a very high
capacity as compared to single-antenna systems. Multiuser MIMO
systems can also exploit most of the advantages of multiple-antenna
systems.

In a broadcast system, when an access point with multiple antennas
is used to communicate with many users, the communication is
complicated by the fact that each user must decode its signal
independently from the others. As a simple precoding scheme, the
channel inversion technique can be used at the transmitter to
separate the data for different users. However, this method is
vulnerable to the poor channel conditions.

In this paper, we investigate the optimum gain for average transmit
energy in multiple antenna broadcast systems with channel inversion
technique. By using the fact that the channel is not orthogonal, the
gain that can be achieved is significantly higher than the regular
shaping gains that can be achieved in methods like \cite{PeHoSw052}.

In a broadcast system with the channel inversion technique (given a
fixed channel matrix), we find the optimal probability distribution
for the data vectors to minimize the average transmit energy. Then,
we introduce a theoretical Selective Mapping (SLM) technique (based
on random coding arguments) to obtain the optimal average transmit
energy. In order to implement the SLM method effectively, we propose
using nested lattice codes in a trellis precoding framework

The rest of the paper is organized as follows. In Section II, the
system model is introduced. Section III finds the optimal
probability distribution for transmit data in channel inversion
techniques. Section IV is devoted to introducing the SLM technique
and its analysis and implementation issues.

\section{System Model}

A multiple antenna broadcast system can be modeled by
\cite{MobKha07}
\begin{eqnarray}\label{eq:realchan}
\mathbf{y} = \mathbf{H} \mathbf{x} + \mathbf{n},
\end{eqnarray}
where $ \mathbf{y} $ is the $\tilde{M}\times 1$ \emph{received
vector}, $ \mathbf{x} $ is the $\tilde{N}\times 1$ \emph{normalized
transmitted data},${\bf n}$ is additive white Gaussian noise, and
${\bf H}$ represents the $\tilde{M}\times \tilde{N}$ channel matrix
in real space.

In broadcast systems, the receivers should decode their respective
data independently and without any cooperation with each other. The
simplest method is using the channel inversion technique as a
precoding method at the transmitter to separate the data for
different users $ \bf{s}=\bf{H}^{+}\bf{u}$, where $
\bf{H}^{+}=\bf{H}^{\ast}(\bf{H}\bf{H}^{\ast})^{-1} $, $
\bf{H}^{\ast}$ is the Hermitian of $\bf{H} $, $ \bf{u} $ is the data
vector, i.e. $ u_{i} $ is the data for the \textit{i}'th user, and $
\bf{s} $ is the transmitted signal before the normalization. When
the number of transmit antennas equals with the number of users,
$\bar{M}=\bar{N}:=M $, the transmitted signal is
\begin{equation}\label{eq:trnsig}
\bf{s}=\bf{H}^{-1}\bf{u}.
\end{equation}
As in \cite{PeHoSw052}, the normalized transmitted signal would be
${\bf x} = \dfrac{\bf{s}} { \sqrt{E \{\gamma\}}}$, where $\gamma =
\| {\bf s} \|^2$. The problem arises when $\bf{H}$ is poorly
conditioned and $ \gamma $ becomes very large, resulting in a high
power consumption.

In a multiple antenna system, it is assumed that the data vector
${\bf u}$ is selected from a constellation with discrete points.
However, through this paper we investigate the probabilistic
behavior of the transmitted signal $\bf{s}$. Assuming a large
constellation, continuous approximation provides a probability
distribution for each constellation, resulting in different
$E\{\gamma\}$. The challenge is finding the best probability
distribution with minimum $E\{\gamma\}$. Note that the expectation
in $E\{\gamma\}$ is over ${\bf u}$ and the channel is assumed
constant.

\section{Optimum Probabilistic Constellation}\label{sec:Prob}

Channel inversion technique removes the need for complex decoding
algorithms in the receiver side; however, it leads to a high energy
consumption as the average energy of the resulting constellation
points is high. We are looking for a constellation shaping method
for the input constellation, such that using the channel inversion
technique, the resulting constellation has a smaller value for the
average transmit energy.

A proper input constellation should be designed such that two
conditions are satisfied: (i) data can be decoded independently at
the receivers, and (ii) the average transmitting energy is as low as
possible. The design of the constellation is known as the shaping.
By using a conventional block constellation, any point in the
constellation is equally likely. However, by shaping, a nonuniform
distribution is achieved over each dimension.

A common constellation shaping technique is to choose a finite set
of points from an $M$ dimensional lattice $\boldsymbol{\Lambda}$
that lies within a finite region $\mathcal{R} \subset \mathds{R}^M$.
This constellation is known as a lattice code. If $\mathbb{C}$ is a
lattice code of reasonably large size, then the distribution of its
points in $M$ dimensional space is well approximated by a uniform
continuous distribution over the region ${\mathcal R}$ (the
continuous approximation) \cite{ForWei89}.

Having a uniform distribution over region ${\mathcal R}$ induces a
nonuniform distribution on each dimension. In other words, if ${\bf
u}$ is selected uniformly over $ \mathcal{R}$, each element of ${\bf
u}$ has a nonuniform distribution. Through this paper, the
probability distribution of the elements of ${\bf u}$ is called
\emph{marginal probability distribution} of ${\bf u}$. We assume
that the region ${\mathcal R}$ has a fixed volume
$\textmd{Vol}(\mathcal{R})= \mathbb{V}$, resulting in the entropy of
($\log \mathbb{V}$). In the case of independent variables, we assume
that the entropy per real dimension is $\mathcal{H}=\frac{1}{M} \log
\mathbb{V}$.

Let ${\bf Q}:=\left(\bf{H}^{-1}\right)^T \bf{H}^{-1}={\bf
U}\boldsymbol{\Lambda}{\bf U}^T$, where $\bf{U}$ is the unitary
matrix of eigenvectors of ${\bf Q}$ and $\boldsymbol{\Lambda}$ is
the diagonal matrix of the corresponding eigenvalues,
$\lambda_i,\;i=1,\cdots,M$. Assume $\bf{u} \in \mathcal{R}$ be a
random vector with mean $E\{{\bf u}\} = \boldsymbol{\mu}$ and the
correlation matrix $E\{{\bf uu}^T\}= \boldsymbol{\Sigma}>0$. The
energy of the transmitted signal, called \emph{transmit energy}, is
defined as $\gamma={\bf u}^T{\bf Qu} = {\bf u}^T{\bf
U}\boldsymbol{\Lambda}{\bf U}^T{\bf u}={\bf v}^T{\bf v}$, where
${\bf v}=\sqrt{\boldsymbol{\Lambda}}{\bf U}^T{\bf u}$. When ${\bf
u}$ is selected uniformly in ${\mathcal R}$, the vector ${\bf v}$ is
selected uniformly over a region $\mathcal{R}^\prime$, where
$\mathcal{R}^\prime=\{ \bf{v} | {\bf v}=\sqrt{\boldsymbol{\Lambda}}
{\bf U}^T{\bf u}, \forall {\bf u}\in \mathcal{R} \}$. It is more
convenient to explain some behaviors of $\gamma$ based on ${\bf v}$.
The average transmit energy can be written as \cite{MobKha07}
\begin{eqnarray}\label{eq:aveE}
E\{\gamma\} = tr ({\bf Q} \boldsymbol{\Sigma}) + \boldsymbol{\mu}^T
\bf{Q} \boldsymbol{\mu}
\end{eqnarray}

If we ignore that users are supposed to decode their data
independent of each other, the optimum region for the input
constellation can be found using the following lemma:

\begin{thm}\label{thm:optGaus}
Let ${\bf u}=[u_1, u_2, \cdots, u_M] \in \mathds{R}^M $ be a random
vector with probability distribution $f(u_1,u_2,\cdots,u_M)$, mean
$E\{{\bf u}\} = \boldsymbol{\mu}$, and the correlation matrix
$E\{{\bf uu}^T\}= \boldsymbol{\Sigma}>0$, in a broadcast system
introduced in (\ref{eq:realchan}). Let $\mathcal{H}({\bf u})$ denote
the entropy of the data vector ${\bf u}$. Then, a multivariate
Gaussian random vector ${\bf u}$ with $\boldsymbol{\mu} = {\bf 0}$
and the covariance matrix
\begin{equation}\label{eq:optSigma}
\boldsymbol{\Sigma}=\sqrt[M]{\Pi\lambda_i} \sigma^2 {\bf HH}^T
\end{equation}
will minimize the energy of the transmit signal given a fixed
entropy $\mathcal{H}({\bf u})=\log(\mathbb{V})$, where $\sigma^2$ is
the variance of a Gaussian random variable with entropy $\mathcal{H}
= \frac{1}{M} \log (\mathbb{V})$.
\end{thm}
\begin{IEEEproof}
See \cite{MobKha07}
\end{IEEEproof}
This choice of $\boldsymbol{\Sigma}$ suggests that the minimum value
of the average energy among transmit signals with different
probability distributions is \cite{MobKha07}
\begin{equation}\label{eq:Eopt}
E_{opt}=E\{\gamma\} = M \sqrt[M] {\Pi\lambda_i}\sigma^2
\end{equation}

Consider the auxiliary vector ${\bf v}= \sqrt{\boldsymbol{\Lambda}}
{\bf Uu}$. It can be easily shown that each element of this vector
has a Gaussian distribution with zero mean with variance
$\mathcal{R}_{eq}^2$. Therefore, in the limit of $M \longrightarrow
\infty$, this vector is uniformly selected over an $M$-dimensional
sphere centered at the origin with radius $\sqrt{M}
\mathcal{R}_{eq}$, i.e. $\mathcal{B}_M(0,\sqrt{M} \mathcal{R}_{eq})$
(corresponding to the minimum average transmit energy in
\eqref{eq:Eopt}.

Roughly speaking, we can assume that the region ${\mathcal
R}^\prime$ is $\mathcal{B}_M(0,\sqrt{M} \mathcal{R}_{eq})$. On the
other hand, the vector ${\bf u}$, a Gaussian random vector with zero
mean and covariance matrix in (\ref{eq:optSigma}), is uniformly
selected over the region ${\mathcal R}$ which is an oval. The main
diameters of this oval are along the eigenvectors ${\bf U}$ and the
radii of the oval in each direction are $\sqrt{\frac{M}
{\lambda_i}}\mathcal{R}_{eq}\textmd{ for }i=1,\cdots,M$, in other
words $\mathcal{R}=\mathcal{O}_M(0,\sqrt{ \frac{M}
{\lambda_i}}\mathcal{R}_{eq})$ . (see \cite{MobKha07}).

By using this region, an additional \emph{channel gain} of
\cite{MobKha07}
\begin{eqnarray}\label{eq:shapgain}
\mathcal{G}_{\bf H} = \dfrac{Arithmetic \; Mean(\lambda_1, \cdots,
\lambda_N)}{Geometric \; Mean(\lambda_1, \cdots, \lambda_N)}
\end{eqnarray}
can be achieved (in addition to the conventional shaping gain).

The geometric mean of a data set is always smaller than or equal to
the set's arithmetic mean (the two means are equal if and only if
all members of the data set are equal). On the other hand, without
the channel matrix, we have the conventional shaping gain. However,
the presence of ${\bf H}^{-1}$ will affect the shaping gain by the
\emph{Channel Gain}, $\mathcal{G}_{\bf H}$, defined in
(\ref{eq:shapgain}). Without the channel effect the optimum region
${\mathcal R}$ is a spherical region (corresponding to independent
Gaussian variables), while with the channel effect the optimum
region ${\mathcal R}$ is an $M$-dimensional oval.

From another point of view, this gain can be seen as the effect of
rate (or power) allocation for Gaussian distribution which has been
considered in multi-carrier transmission and point to point multiple
antenna systems, e.g. \cite{FisHub97}. However, this concept ignores
the independency condition required for a broadcast system. Here,
the challenging problem is how the region $\mathcal{R}$ or
$\mathcal{R}^\prime$ can be achieved, while considering the
independency condition.

\section{Selective Mapping}

The idea of Selective Mapping (SLM) is to generate a large set of
data vectors that represent the same information, where the data
vector resulting in the lowest energy is selected for transmission.
This idea has been used in OFDM systems, e.g. \cite{MobKha06}, to
reduce the average transmit energy.

In the optimum case, the vector ${\bf u}$ is selected uniformly over
an $M$-dimensional oval and the transmit vector is selected
uniformly over a hypersphere. However, due to the independency
condition, implementing this oval shape region is not possible. The
receivers can not co-operate with each other to locate a point
inside this oval. We propose an SLM method that can theoretically
achieve the optimum gain for average transmit energy. The region for
vector ${\bf u}$ is not oval; however, the resulting region for the
transmit vector in the limit is a hypersphere.

In the sequel, first, we use a random coding argument to explain the
SLM method, its analysis, and the maximum theoretical gain that can
be achieved. In this part, again we ignore the independency
condition. In continue, we implement the SLM technique considering
the independency condition by using a trellis precoding.

In the system model (\ref{eq:realchan}), the volume of the region is
fixed, ${\mathbb V}$. In order to provide multiple choices for the
SLM method, the volume is increased to $\bar{\mathbb V}$ such that
for each data vector there are $N$ points, where $ N =
\dfrac{\bar{\mathbb V}}{\mathbb V}$.

In other words, $N$ i.i.d. samples of ${\bf u}$ are generated, $\{
{\bf u}_1, {\bf u}_2, \cdots, {\bf u}_{N} \}$, and ${\bf s}_l$ with
the lowest transmit energy is selected for transmission. In other
words, $ \gamma_l = \min \{ \gamma_1, \gamma_2, \cdots, \gamma_{N}
\}$. We are looking for the probabilistic behavior of $\gamma_l$.

\subsection{Asymptotic Analysis}

In this section, we analysis the effect of an SLM method for
broadcast systems. In the proposed method, $N$ i.i.d. samples of
${\bf u}$ are generated, $\{ {\bf u}_1, {\bf u}_2, \cdots, {\bf
u}_{N} \}$, and among the corresponding transmit vectors ${\bf s}_i
= {\bf H}^{-1}{\bf u}_i$, the vector ${\bf s}_l$ with the lowest
transmit energy is selected for transmission. In other words, in the
SLM method, we are looking for
\begin{equation}\label{eq:slmmin}
\min_{1 \leq i \leq N} \|{\bf s}_i\|^2,
\end{equation}
where $\|.\|$ represents the regular norm.

The expression in (\ref{eq:slmmin}) is similar to minimization of
distortion in quantization and random quantizers. The tremendous
research on random quantization \cite[and ref. therein]{GraLus00}
can help us to evaluate the expression in (\ref{eq:slmmin}) in our
SLM method.

Let ${\bf s}_1, {\bf s}_2, \cdots, {\bf s}_{N}$, be i.i.d.
$\mathds{R}^M$-valued random variables with distribution $Q$, i.e.
\begin{equation}
Q({\bf v}) = \mathds{P} \{ s_{i_1} \leq v_1, \cdots , s_{i_M} \leq
v_M \} \quad i=1,\cdots,N,
\end{equation}
where
$${\bf v}=(v_1, \cdots,v_M) \in \mathds{R}^M.$$
For any region $\mathcal{R}$, the probability $Q(\mathcal{R})$ is
the probability that there is at least one code point in the region
$\mathcal{R}$, i.e.
$$Q(\mathcal{R})=\int_{\mathcal{R}} Q(d{\bf y}).$$

Define the $r^{th}$ \emph{order transmit energy} as
\begin{equation}\label{eq:renergy}
\gamma_{r, N}^{Q} = \min_{1\leq i \leq N} \|{\bf s}_i\|^r,
\end{equation}
where based on our previous notation $\gamma_l = \gamma_{2,N}^{Q}$.
In this section, the asymptotic probabilistic behavior of
$\gamma_{2,N}^{Q}$, when $N\longrightarrow \infty$, is investigated.
Specifically, we calculate the average transmit energy in the SLM
technique. Note that, in the following, we frequently use $\lambda$
which is the $M$-dimensional Lebesgue measure. Here, we define it as
the $M$-dimensional volume of a region.

\begin{thm}\label{thm:asymp}
Let ${\bf s}_1, {\bf s}_2, \cdots, {\bf s}_{N}$, be i.i.d.
$\mathds{R}^M$-valued random variables with distribution $Q$. Then,
\begin{equation}\label{eq:avegen}
\lim_{N\rightarrow\infty} E\left\{ N^{\frac{r}{M}}\gamma_{r, N}^{Q}
\right\} =
B_M^{-\frac{r}{M}}\Gamma(1+\frac{r}{M})g_\rho^{-\frac{r}{M}}
\end{equation}
where $B_1=2$, $B_M=\lambda \left( \mathcal{B}_M(0, 1) \right) =
\pi^{M/2}/ \Gamma(1+M/2)$ for $M=2,\cdots$, and $g_\rho$ is defined
for any $\rho>0$ as
\begin{displaymath}
g_\rho := \inf_{\delta \in (0,\rho]} \dfrac{Q\left( \mathcal{B}_M(0,
\delta) \right)}{\lambda\left( \mathcal{B}_M(0, \delta) \right)}.
\end{displaymath}
\end{thm}

\begin{IEEEproof}
See \cite{MobKha07}.
\end{IEEEproof}

Now, consider the special case of uniform distribution. When we have
a large lattice code, we can assume we have a uniform distribution
over the region where the lattice code is defined. Applying SLM
technique, over a region with uniform distribution results in the
following average for the $r^{th}$ order transmit energy.
\begin{thm}\label{thm:aveuniform}
Let $\mathcal{R} \subset \mathds{R}^M$ be a compact set with
$\lambda(\mathcal{R})>0$ and let ${\bf s}_1,\cdots,{\bf s}_N$ be
i.i.d. random variables with uniform distribution over
$\mathcal{R}$. Then,
\begin{equation}\label{eq:limSLM}
\lim_{N\rightarrow\infty} E\left\{ N^{\frac{r}{M}}\gamma_{r, N}^{Q}
\right\} = B_M^{-\frac{r}{M}} \Gamma(1+\frac{r}{M})
\lambda(\mathcal{R})^{\frac{r}{M}}.
\end{equation}
\end{thm}
\begin{IEEEproof}
Let $Q$ be a uniform distribution over $\mathcal{R}$, i.e.
$Q=U(\mathcal{R})$. Therefore,
\begin{equation}
Q\left( \mathcal{B}_M(0, \dfrac{v^{\frac{1}{r}}}{N^{\frac{1}{M}}})
\right) = \dfrac{\lambda\left(\mathcal{B}_M(0,
\dfrac{v^{\frac{1}{r}}}{N^{\frac{1}{M}}})\right)}{\lambda(\mathcal{R})},
\end{equation}
and
\begin{equation}\label{eq:guniform}
g_\rho = \inf_{\delta \in (0,\rho]} \dfrac{Q\left( \mathcal{B}_M(0,
\delta) \right)}{\lambda\left( \mathcal{B}_M(0, \delta)
\right)}=\dfrac{1}{\lambda(\mathcal{R})}.
\end{equation}
Substituting (\ref{eq:guniform}) in (\ref{eq:avegen}) completes the
proof.
\end{IEEEproof}

Note that we are interested in cases that the i.i.d random variables
${\bf u}_1, \cdots, {\bf u}_N$ are selected uniformly over a region
$\mathcal{R}^\prime$. According to ${\bf s}={\bf H}^{-1}{\bf u}$,
for the probability distribution of ${\bf s}$, we have
\begin{equation}
f_{\bf s}({\bf s})=|{\bf H}^{-1}|\;f_{\bf u}({\bf Hs}).
\end{equation}
Therefore, if ${\bf u}$ has a uniform distribution over
$\mathcal{R}^\prime$, ${\bf s}$ has also a uniform distribution over
$\mathcal{R}$, where $\mathcal{R} = {\bf H}^{-1}
\mathcal{R}^\prime$.

In order to find the asymptotic average transmit energy of SLM
technique, we should replace $r=2$ and $Q=U(\mathcal{R})$, where
$\mathcal{R}$ is the region for the transmit vector ${\bf s}$.
Therefore, according to the expression in (\ref{eq:limSLM}), the
average transmit energy for large $N$ can be approximated by
\cite{MobKha07}
\begin{equation}\label{eq:ESLM}
E_{SLM}=\Gamma(1+\frac{2}{M}) M \mathcal{R}_{eq}^2.
\end{equation}
Comparing \eqref{eq:Eopt} and \eqref{eq:ESLM}, we can see that using
SLM technique with any lattice code of reasonably large size the
optimum transmit energy can be achieved since for large $M$,
$\Gamma(1+\frac{2}{M})\longrightarrow\Gamma(1)=1$.
\begin{col}
In a broadcast system, applying SLM method to lattice codes of
reasonably large size, with a fixed volume, will result in equal
values for the average transmit energy when $N$ is large enough in
the SLM method.
\end{col}

We must emphasis that in our random coding argument the probability
of the event that two different code words have the same transmit
data vector is negligible. In the case of this event, we have an
error in our broadcast system. However, since the probability of
this event is small, the average transmit energy would not change.

\subsection{Implementation Issues}

In any practical SLM method, the lattice code $\mathbb{C}$
(constellation) should be expanded such that the number of
constellation points are multiplied by $N$, resulting in a new
lattice code $\mathbb{C}^\prime$. This new set of constellation
points are grouped in $|\mathbb{C}|$ sets containing $N$ points.
Transmitting any of these $N$ points transfer the same information.
These sets (and the expanded constellation) should be selected such
that the users at the receive side can decode their data independent
of each other.

The method proposed in \cite{PeHoSw052} can be considered as an SLM
technique with this idea. In this method the region for the transmit
vector ${\bf s}$ is expanded by repetition of the constellation by
multiples of $\tau$ in each direction. In other words, for any
vector ${\bf s}={\bf H}^{-1}{\bf u}$, we find ${\bf H}^{-1}({\bf u}
+ \tau {\bf l})$ for $\lfloor -b/2 \rfloor+1 \leq l_i \leq \lfloor
b/2 \rfloor$. In each direction we repeat the constellation $b$
times, so we have $N=b^M$ in the SLM method. In this method, the
transmit vector ${\bf s}$ is selected in the original constellation
and $N-1$ other points are calculated by adding integer vector
offsets, resulting in $N$ points in the expanded lattice code
$\mathbb{C}^\prime$. A modulo operation in the transmitter and
receivers guarantees and satisfies the independency condition.

For large enough $M$ and $N$, this method can not achieve the
optimum average energy. The equivalent region for vector ${\bf s}$
is the Voronoi region of $\tau {\bf H}^{-1}$, not a hypersphere
\cite{MobKha07}. This leads to an improvement over the Gaussian
marginal probability distribution; however, this is not the best
that we can achieve. The more this Voronoi region looks like a
sphere, the less the average transmit energy is. The problem in the
SLM method in \cite{PeHoSw052} is that the vector ${\bf s}$ is not
uniformly distributed over lattice code ${\mathbb C}^\prime$. In
order to preserve the independency condition, a vector is uniformly
distributed over $\mathbb{C}$ and $N-1$ other points are calculated
deterministically in ${\mathbb C}^\prime$ based on this point. This
results in a region with Voronoi region shape not a sphere.

In \cite{CaKhSa07}, a sign-bit shaping algorithm is proposed for
precoding in broadcast systems. Sign-bit shaping is implemented by
using a trellis code. This technique is actually an SLM method since
it gives the transmitter many different options when determining
which symbol to transmit.

Trellis shaping systems are composed of a rate $(k_s,n_s)$ binary
convolutional shaping code $C$ and a signal set $A$ partitioned into
$2^{n_s}$ shaping subsets \cite{EyuFor92}. The signal set A is
typically a lattice code with shaping region $\mathcal{R}$, and the
shaping subsets are the points of this region that fall within
subregions $\mathcal{R}_i$ for $i=1,\cdots,2^{n_s}$. It is important
that the code $C$ and the signal set $A$ is selected such that the
the equivalent points are selected uniformly over $\mathcal{R}$.

Conventional precoding schemes in broadcast systems, such as
\cite{PeHoSw052}, treat multiple antennas of different users as
different users. By using trellis shaping for each virtual user, in
each 2-dimensional space, there is a modulo operation with respect
to the Voronoi region of the shaping trellis code. In other words,
in the space of each user, there is a modulo operation with respect
to the Cartesian product of these Voronoi Regions. However, we can
use the shaping concept in each user's space. The idea in
\cite{CaKhSa07} can be extended to include the multi-antenna case.
We can use a trellis shaping for each user, and not for each
antenna.

Nested lattice codes can be implemented such that both these
improvements are met. The idea of nested lattice codes has already
been used for interference cancelation in degenerated broadcast
systems \cite{ZaShEr02}. There, it is assumed that, in an ordered
set of users, each user has the ability that it can decode the
message for the previous users. In other words, it is assumed that
each user has the code-book for the previous users. This technique
can be implemented in our scheme to provide us with shaping, without
any need for these assumptions. We can achieve the same gain as that
reported in \cite{ZaShEr02} for broadcast systems with precoding,
without any extra assumption.

Assume that in a broadcast system with $K$ users, each user has
$n_u$ antenna in \eqref{eq:realchan}, i.e. $M= 2Kn_u$. One way of
implementing this idea is implementing a large trellis consisting of
$K$ sub-trellises with a lattice partition, $\boldsymbol{\Lambda}/
\boldsymbol{\Lambda}^\prime$, in a $2n_u$ dimensional space. In each
sub-trellis, the lattice code is divided into
$|\boldsymbol{\Lambda}/ \boldsymbol{\Lambda}^\prime|$ partitions,
i.e. for transmitting any information vector, one of
$|\boldsymbol{\Lambda}/ \boldsymbol{\Lambda}^\prime|^K$ equivalent
points, in $M$ dimensional space, with the lowest transmit energy is
selected. In other words, the vector ${\bf u}$ resulting in the
lowest energy for ${\bf H}^{-1}{\bf u}$ would be selected for
transmission. Now, in each $2n_u$ dimensional space, the modulo
operation is with respect to the Voronoi region of this trellis
code. The Cartesian product of these regions should be as close as
possible to the sphere in order to generate the optimum shaping
region.

%
%

\end{document}